\documentclass[onecolumn,superscriptaddress,aps,prl,preprint,showkeys]{revtex4-1}
\usepackage{color}
\usepackage{hyperref}
\usepackage{amssymb}
\usepackage{amsmath}

\hypersetup{colorlinks,linkcolor=blue,citecolor=blue,urlcolor=blue}

\usepackage{mhchem}
\usepackage{latexsym}
\usepackage[pdftex]{graphicx} %zum einbinden von xfig grafiken
\usepackage{xspace}

\begin{document}
	
	\title{Probing edge state conductance in ultra-thin topological insulator films}
	
	%\today
	
	\author{Arthur Leis}
	%\email{a.leis@fz-juelich.de}
	\affiliation{Peter Gr\"{u}nberg Institut (PGI-3), Forschungszentrum J\"{u}lich, 52425 J\"{u}lich, Germany}
	
	\affiliation{J\"ulich Aachen Research Alliance (JARA), Fundamentals of Future Information Technology, 52425 J\"ulich, Germany}
	\affiliation{Experimentalphysik IV A, RWTH Aachen University, Otto-Blumenthal-Stra\ss{}e, 52074 Aachen, Germany}
	
	\author{Michael Schleenvoigt}
	%\email{m.schleenvoigt@fz-juelich.de}
	\affiliation{J\"ulich Aachen Research Alliance (JARA), Fundamentals of Future Information Technology, 52425 J\"ulich, Germany}
	\affiliation{Peter Gr\"{u}nberg Institut (PGI-9), Forschungszentrum J\"{u}lich, 52425 J\"{u}lich, Germany}
	
	\author{Kristof Moors}
	%\email{k.moors@fz-juelich.de}
	\affiliation{Peter Gr\"{u}nberg Institut (PGI-9), Forschungszentrum J\"{u}lich, 52425 J\"{u}lich, Germany}
	
	\author{Helmut Soltner}
	%\email{h.soltner@fz-juelich.de}
	\affiliation{Zentralinstitut f\"{u}r Engineering, Elektronik und Analytik (ZEA-1), Forschungszentrum J\"{u}lich, 52425 J\"{u}lich, Germany}
	
	\author{Vasily Cherepanov}
	%\email{v.cherepanov@fz-juelich.de}
	\affiliation{Peter Gr\"{u}nberg Institut (PGI-3), Forschungszentrum J\"{u}lich, 52425 J\"{u}lich, Germany}
	\affiliation{J\"ulich Aachen Research Alliance (JARA), Fundamentals of Future Information Technology, 52425 J\"ulich, Germany}
	
	\author{Peter Sch\"uffelgen}
	%\email{p.schueffelgen@fz-juelich.de}
	\affiliation{J\"ulich Aachen Research Alliance (JARA), Fundamentals of Future Information Technology, 52425 J\"ulich, Germany}
	\affiliation{Peter Gr\"{u}nberg Institut (PGI-9), Forschungszentrum J\"{u}lich, 52425 J\"{u}lich, Germany}
	
	\author{Gregor Mussler}
	%\email{g.mussler@fz-juelich.de}
	\affiliation{J\"ulich Aachen Research Alliance (JARA), Fundamentals of Future Information Technology, 52425 J\"ulich, Germany}
	\affiliation{Peter Gr\"{u}nberg Institut (PGI-9), Forschungszentrum J\"{u}lich, 52425 J\"{u}lich, Germany}
	
	\author{Detlev Gr\"utzmacher}
	%\email{d.gruetzmacher@fz-juelich.de}
	\affiliation{J\"ulich Aachen Research Alliance (JARA), Fundamentals of Future Information Technology, 52425 J\"ulich, Germany}
	\affiliation{Peter Gr\"{u}nberg Institut (PGI-9), Forschungszentrum J\"{u}lich, 52425 J\"{u}lich, Germany}
	
	\author{Bert Voigtl\"ander}
	\email{b.voigtlaender@fz-juelich.de}
	\affiliation{Peter Gr\"{u}nberg Institut (PGI-3), Forschungszentrum J\"{u}lich, 52425 J\"{u}lich, Germany}
	\affiliation{J\"ulich Aachen Research Alliance (JARA), Fundamentals of Future Information Technology, 52425 J\"ulich, Germany}
	\affiliation{Experimentalphysik IV A, RWTH Aachen University, Otto-Blumenthal-Stra\ss{}e, 52074 Aachen, Germany}
	
	\author{Felix L\"upke}
	%\email{f.luepke@fz-juelich.de}
	\affiliation{Peter Gr\"{u}nberg Institut (PGI-3), Forschungszentrum J\"{u}lich, 52425 J\"{u}lich, Germany}
	\affiliation{J\"ulich Aachen Research Alliance (JARA), Fundamentals of Future Information Technology, 52425 J\"ulich, Germany}
	
	\author{F. Stefan Tautz}
	%\email{s.tautz@fz-juelich.de}
	\affiliation{Peter Gr\"{u}nberg Institut (PGI-3), Forschungszentrum J\"{u}lich, 52425 J\"{u}lich, Germany}
	\affiliation{J\"ulich Aachen Research Alliance (JARA), Fundamentals of Future Information Technology, 52425 J\"ulich, Germany}
	\affiliation{Experimentalphysik IV A, RWTH Aachen University, Otto-Blumenthal-Stra\ss{}e, 52074 Aachen, Germany}

	\pacs{}
	\keywords{topological insulator, quantum spin Hall effect, multi-tip scanning tunneling microscopy, nanoscale charge transport, ballistic conductivity, edge channel}
	
	\begin{abstract}
		\textbf{ 
			Quantum spin Hall (QSH) insulators have unique electronic properties, comprising a band gap in their two-dimensional interior and one-dimensional spin-polarized edge states in which current flows ballistically.  In scanning tunneling microscopy (STM), the edge states manifest themselves as a localized density of states. However, there is a significant research gap between the observation of edge states in nanoscale spectroscopy, and the detection of ballistic transport in edge channels which typically relies on transport experiments with microscale lithographic contacts. Here, we study few-layer films of the three-dimensional topological insulator (Bi$_{x}$Sb$_{1-x})_2$Te$_3$, for which a topological transition to a two-dimensional topological QSH insulator phase has been proposed. Indeed, an edge state in the local density of states is observed within the band gap. Yet, in nanoscale transport experiments with a four-tip STM, 2 and 3 quintuple layer films do not exhibit a ballistic conductance in the edge channels. This demonstrates that the detection of edge states in spectroscopy can be misleading with regard to the identification of a QSH phase. In contrast, nanoscale multi-tip transport experiments are a robust method for effectively pinpointing ballistic edge channels, as opposed to trivial edge states, in quantum materials.}
	\end{abstract}

	\maketitle
	
	\section{Introduction}
	A quantum spin Hall (QSH) phase is characterized by a band gap in its two-dimensional interior and by one-dimensional, counter-propagating and spin-polarized helical edge states \cite{Bernevig2006,Lodge2021}. Inducing superconductivity in QSH edge states provides routes to engineering topological superconductivity, with potential applications in topological quantum computing \cite{Sarma2015,Sato2017}.
	% \cite{Kitaev2001,Fu2008,Sarma2015,Sato2017}
	So far, the QSH phase has been realized, e.g., in semiconductor quantum wells \cite{Koenig2007} and monolayer transition metal dichalcogenides \cite{Tang2017, Fei2017, Wu2018}. The unequivocal identification of the QSH phase in a given material requires the confirmation of two experimental signatures: an electronic band gap in the  bands of two-dimensional interior, and spin-polarized helical edge states with a ballistic conductance around the perimeter of the topological phase, i.e., at the boundaries to topologically trivial matter or vacuum.
	
\begin{figure}
		\includegraphics[width=0.60\linewidth]{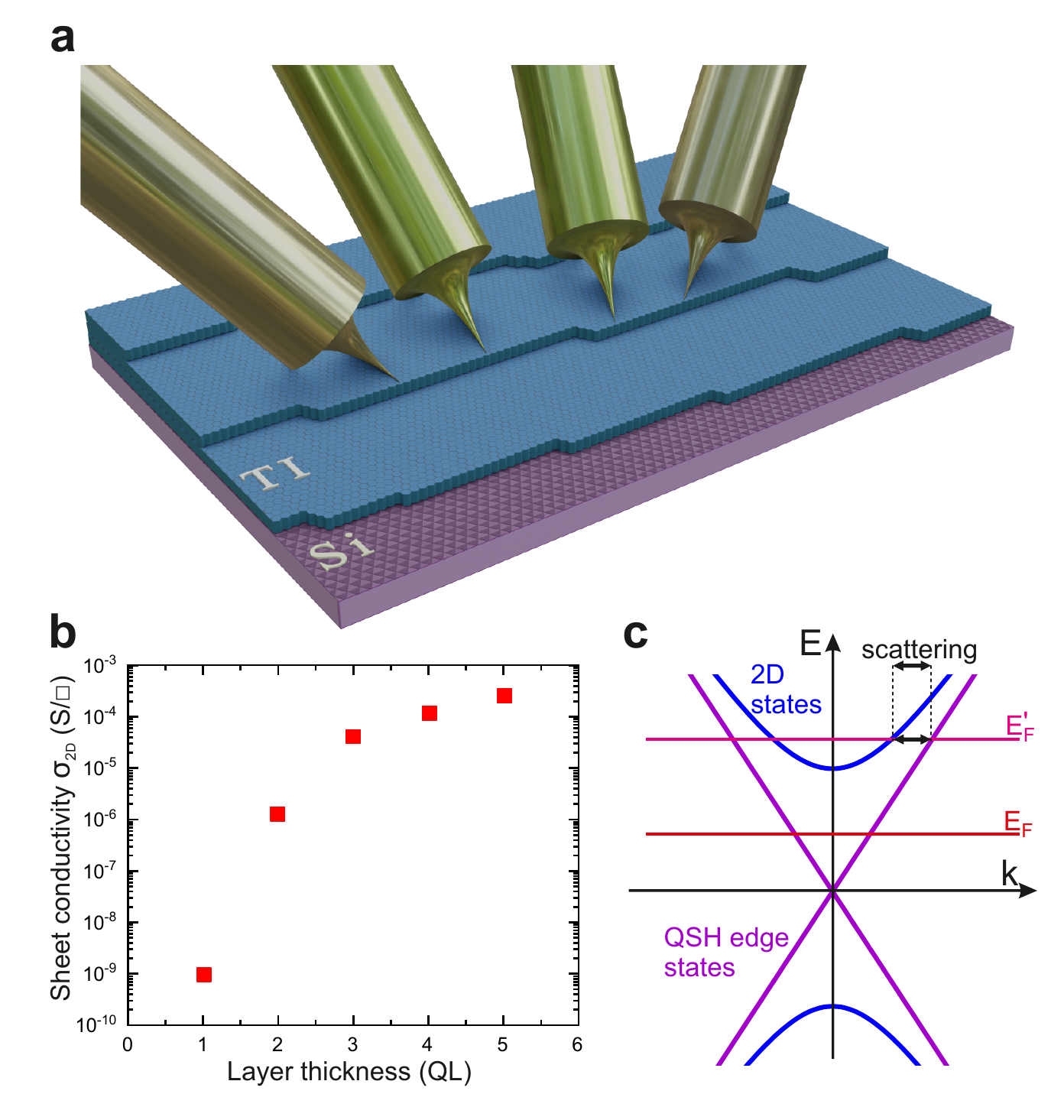}
		\caption{Schematic sketch of the sample layout and the principle of non-invasive nanoscale transport measurements with freely positionable contacts. a) Due to the shadow mask technique employed during the MBE growth of the sample, the boundary region of the TI film forms a wedge, consisting of terraces terminated by single-layer step edges. Positioning the tips of the four-tip STM parallel or perpendicular to the step edges and using them as electrical probes allows transport measurements involving possible edge states. Since the two-dimensional sheet conductivities $\sigma_\mathrm{2D}$ of each terrace are known, the contribution of the edge states to the overall charge transport can be identified. b) Terrace sheet conductivities $\sigma_\mathrm{2D}$ of (Bi$_{0.16}$Sb$_{0.84})_2$Te$_3$ thin films as a function of thickness, measured in quintuple layers (QL). Data are reproduced from reference \cite{Leis2021}. c) Schematic band structure of a two-dimensional TI film. Due to the interaction between top and bottom TSS of the underlying three-dimensional TI, a band gap forms in the two-dimensional interior of the film; corresponding states are shown in blue. At the edges of the two-dimensional TI, topologically protected QSH edge states may form (purple). If the Fermi level is located in the gap ($E_F$), the transport in them is expected to be ballistic, while the film interior shows a comparatively low sheet conductivity $\sigma_\mathrm{2D}$. If the Fermi level is located outside the gap ($E'_F$), $\sigma_\mathrm{2D}$ is larger and inter-band scattering may destroy the ballistic conductivity in the QSH edge states. 
		}
		\label{fig:fig1}
	\end{figure}

	Although local probes, such as scanning tunneling spectroscopy (STS), are powerful tools to characterize the bulk band gap and the localized densities of states (DOS) of possible edge states \cite{Jiang2012}, such measurements cannot provide sufficient evidence for a QSH phase. The localized DOS of an edge state is at most a necessary, but certainly not a sufficient condition for its existence. In contrast, a ballistic edge conductivity constitutes a much more robust signature of a QSH phase, if it can be measured reliably. The latter is indeed challenging, because ultra-thin films are difficult to prepare with well-defined edges, and moreover they are sensitive to degradation in ambient conditions and lithographic processing. These adversities can distort or even mask the QSH effect, especially if the transport measurements provide no information on the spatial distribution of the electrical current---a common disadvantage of standard electrical transport measurements with lithographic contacts. A clearcut proof of a QSH phase in real systems is therefore facilitated considerably by a combination of nanoscale imaging, local spectroscopy and local transport characterization, ideally under in-situ conditions. This unique portfolio is only offered by multi-tip scanning tunneling microscopy (STM), in which each tip serves as a mobile electrical probe \cite{Voigtlaender2018}, as shown in Fig.~\ref{fig:fig1}a. 
	
	Whereas the QSH phase was first observed in HgTe quantum wells \cite{Koenig2007}, the weakly coupled nature of the layered van der Waals material (Bi$_{x}$Sb$_{1-x}$)$_2$Te$_3$ (BST) allows direct access to the topological surface states for studying their detailed properties, and BST also allows a more flexible integration of the QSH phase into devices with other materials. Here, we use multi-tip STM to study edge state conductances in few-layer BST films. In the bulk limit, BST is a three-dimensional topological insulator (TI) with two-dimensional topological surface states (TSS) and a bulk band gap of $\sim300\,$meV \cite{Zhang2009, Xia2009, Chen2009, Hsieh2009}. However, when thinned down to a few quintuple layers (QL), the TSS at the top and bottom surfaces of the film begin to hybridize, resulting in a topological transition to either a QSH or a trivial insulator, depending on the material composition \cite{Lu2010, Liu2010, Kim2012, Foerster2015, Foerster2016}. Both topological phase transitions are accompanied by the opening of a gap at the surface-state Dirac points, but in contrast to the topologically trivial phase the QSH phase guarantees the presence of topological edge states and corresponding ballistic conductance channels. 
	
	The opening of a gap at the Dirac point of BST in the few-QL limit was observed by both by photoemission spectroscopy and STS \cite{Zhang2010, Jiang2012, Stroscio2013}. Moreover, four-point resistance measurements on individual terraces of a BST sample showed an approximately exponential drop of the sheet conductivity with decreasing film thickness \cite{Leis2021}, as displayed in Fig.~\ref{fig:fig1}b. But the observation of the hallmark of a QSH phase, i.e., a ballistic conductance in a topological edge state, has not yet been confirmed in this material system. Due to the required lithographic patterning of electrical contacts, which results in the immediate degradation of the ultra-thin films \cite{Ngabonziza2015}, such measurements are very challenging in the standard methodology of transport experiments. With our multi-tip STM approach, however, we circumvent this problem elegantly and are able to access the electrical properties of the pristine edge states in situ, without the need for lithographic processing.
	
	We have chosen a Sb-rich ternary BST compound, (Bi$_{0.16}$Sb$_{0.84}$)$_2$Te$_3$, for our search of highly conductive edge states, because we found before that close to this stoichiometry the intrinsic $n$ doping of Bi$_2$Te$_3$ and the intrinsic $p$ doping of Sb$_2$Te$_3$ balance each other out \cite{Zhang2011}. It is well-known that the potential occurrence of a QSH phase in ultra-thin films depends both on the composition and the thickness of the film. For binary BST compounds, theoretical calculations predicted that only the 2\,QL Bi$_2$Te$_3$ film supports a QSH phase \cite{Foerster2016}, while other calculations predicted 3\,QL Sb$_2$Te$_3$ to form a QSH phase under specific conditions \cite{Kim2012}. Also, some experimental indication of a QSH phase in 3\,QL Sb$_2$Te$_3$ was found \cite{Stroscio2013}. In our experiments we confirm that 2\,QL and 3\,QL films with the composition (Bi$_{0.16}$Sb$_{0.84}$)$_2$Te$_3$ do not exhibit highly conductive edge states. For 2\,QL, we conclude that the film does not support a QSH phase. Notably, the absence of conducting edge states in our transport measurements occurs in spite of the spectroscopic identification of edge states, which therefore must be assigned to trivial ones, not exhibiting ballistic transport. This illustrates the value of our generic approach to the unambiguous pinpointing of topological edge states in quantum materials, although for the specific sample under study no such states could be identified.  
	
	\section{Experiments}
	(Bi$_{0.16}$Sb$_{0.84}$)$_2$Te$_3$ thin films were grown on silicon-on-insulator (SOI) substrates by molecular beam epitaxy (MBE), using the same process as reported earlier \cite{Leis2021}. The use of SOI substrates with a thin top layer of intrinsic Si reduces the substrate sheet conductivity to $\sim2 \,$nS$/\square$. To achieve a TI film with boundaries on the SOI substrate without the need of ex situ processing, growth was conducted through a removable shadow mask.  In the boundary region of the latter, the TI film  formed a wedge in which the thickness decreases in single-QL steps  from the maximum  of 12\,QL ($1\, \mathrm{QL} \approx 1 \,$nm)  down to the bare Si(111) template layer, as indicated in Fig.~\ref{fig:fig1}a. After the growth, vacuum transfer with $p\leq1\cdot10^{-9}\rm\,mbar$ was carried out to load the sample into the room temperature four-tip STM ($p\leq4\cdot10^{-10}\rm\,mbar$).  The vacuum transfer, as well as the fact that electrical transport measurements using a four-tip STM do not require any additional sample processing beyond the growth, both preserve the pristine TI film, thus avoiding any influence of passivation or lithography steps on the charge transport properties \cite{Ngabonziza2015}. After the electrical measurements, ex situ Rutherford backscattering measurements were performed to determine the precise atomic composition of the ternary compound.
	
	All STM, STS and nanoscale four-point transport measurements reported here were performed in the Jülich room-temperature multi-tip STM \cite{Cherepanov2012}. Topographic STM images and $\mathrm{d}I/\mathrm{d}V$-images were recorded in constant-current mode with bias voltage $V_\mathrm{b}$ applied to the sample. The spectroscopic $\mathrm{d}I/\mathrm{d}V$ signal was acquired using standard lock-in techniques with a modulation frequency $f= 320\,$Hz and amplitude $V_\mathrm{mod} = 30\,$mV. To first order, $\mathrm{d}I/\mathrm{d}V(E)$ is proportional to the local density of states (LDOS) at the local probe position and at energy $E = eV_\mathrm{b}$ with respect to the Fermi level of the sample \cite{SPMbook}. Electrochemically etched tungsten wires were used as STM tips.
	
	To perform electrical four-point measurements with the multi-tip STM, the four tips were brought to the desired positions on the sample surface by scanning the region of interest in tunneling contact. They were then lowered from the tunneling regime into contact with the sample surface, thereby establishing electrical point contacts to the TI film \cite{Leis2020}. After the electrical measurements the tips were lifted from the sample surface and subsequent topography scans showed that the contact points of the tips are discernible only as small spots of $\sim 1 \,$nm height, showing that the point contacts are nearly non-invasive and not influencing the measured four-point resistances. 
	
\begin{figure}
		\includegraphics[width=\linewidth]{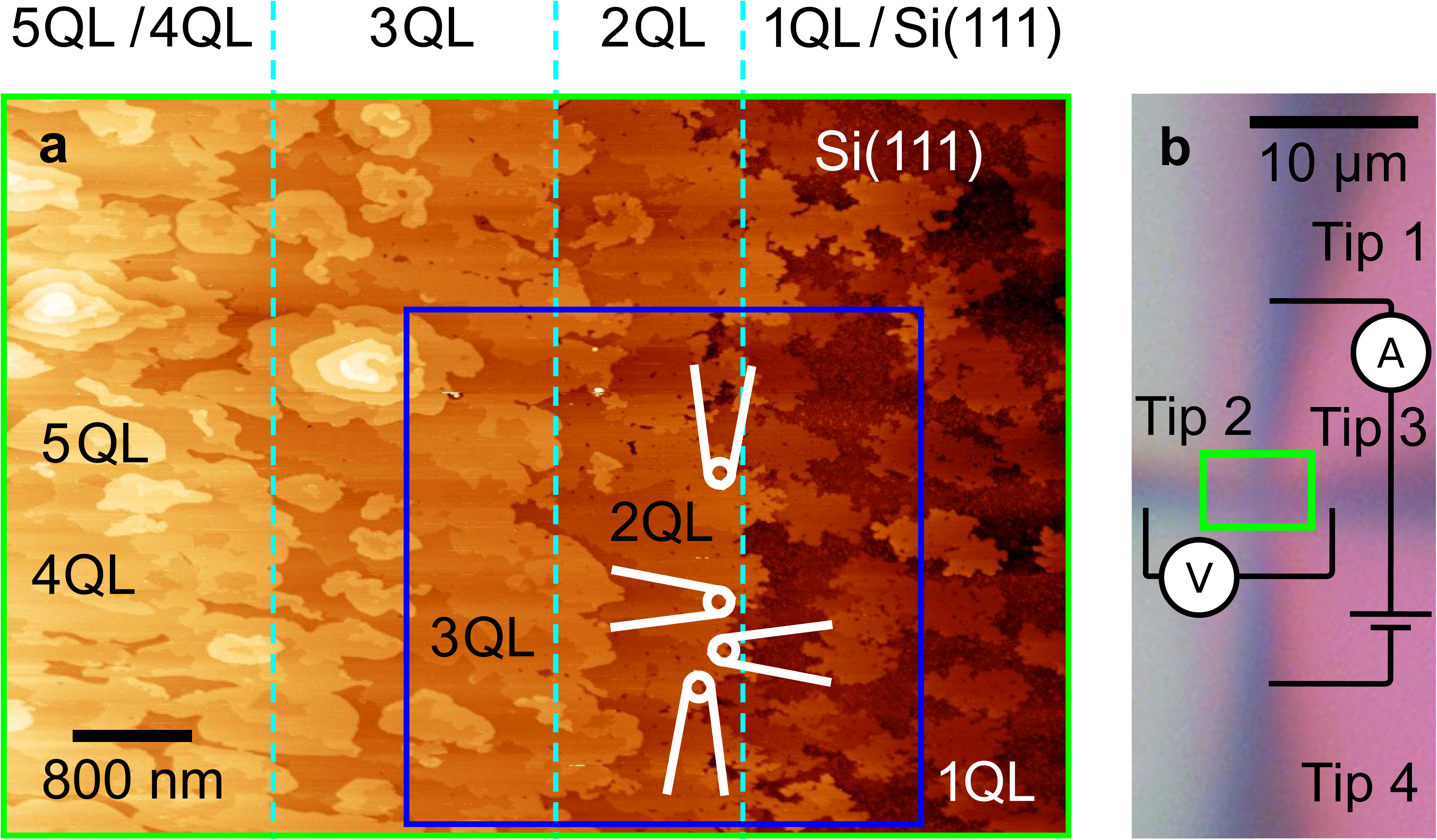}
		\caption{Topography of the (Bi$_{0.16}$Sb$_{0.84})_2$Te$_3$ film and exemplary measurement configuration. a) A large overview STM scan is performed with one of the tips to map the topography of the TI film close to its boundary. The image shows single-QL steps (cyan dashed lines), thus revealing the wedge-shaped structure of the film down to the Si(111) substrate. Film thicknesses are labeled above the map. The overview scan serves as a reference map to place all four tips (white symbols) in the vicinity of a selected step edge (here: 1\,QL\,/\,2\,QL). The blue square indicates the image frame of Fig.~\ref{fig:fig4}(a).  b) Optical microscope image of the final tip configuration. The area of the overview STM scan is indicated by the green rectangle.}
		\label{fig:fig2}
	\end{figure}

	In the present experiments, placing the tips close to the step edges of the MBE-grown wedge-like film is a particular challenge. To meet the required level of precision in placing the tips close to a selected step edge, we employ a positioning technique which relies on successive STM scans performed with all tips \cite{Leis2020, Leis2022}. In a first step, we use the optical microscope to position the tips roughly above the boundary region (Fig.~\ref{fig:fig2}b). Then, we approach the tips into the tunneling regime and measure overview scans of the topography of the TI film. Figure \ref{fig:fig2}a shows such a large overview scan performed with one of the tips in the boundary region of the film, indicating the approximate extent of the terraces by the dashed blue lines. After moving all tips into the mapped boundary region (green rectangle in Fig. \ref{fig:fig2}b), using optical microscope imaging, small-scale STM scans are recorded with each tip to find topographic features that are also seen in the large overview scan. Once a topographic feature is recognized, the position of the corresponding scanning tip is known. Further adjustments of each tip position can then be achieved by fine lateral movement using piezoelectric control in tunneling contact \cite{Leis2022}. In this manner, all tips are placed in a desired position, as shown exemplary by the white symbols in Fig. \ref{fig:fig2}a.
	
	In the ensuing transport experiment, the outer two STM tips inject a charge current, while the inner tips probe the resulting electrochemical potential. The four-point resistance for a given tip arrangement is measured by recording the four-probe $I$-$V$ characteristics, with currents of up to 10\,$\mathrm{\mu A}$. In all measurements, only the position of one of the voltage-probing tips is varied, while the other tips are kept in fixed locations.

	\section{Results and Discussion}
	\subsection{Spectroscopic evidence for edge states}
\begin{figure}
		\includegraphics[width=\linewidth]{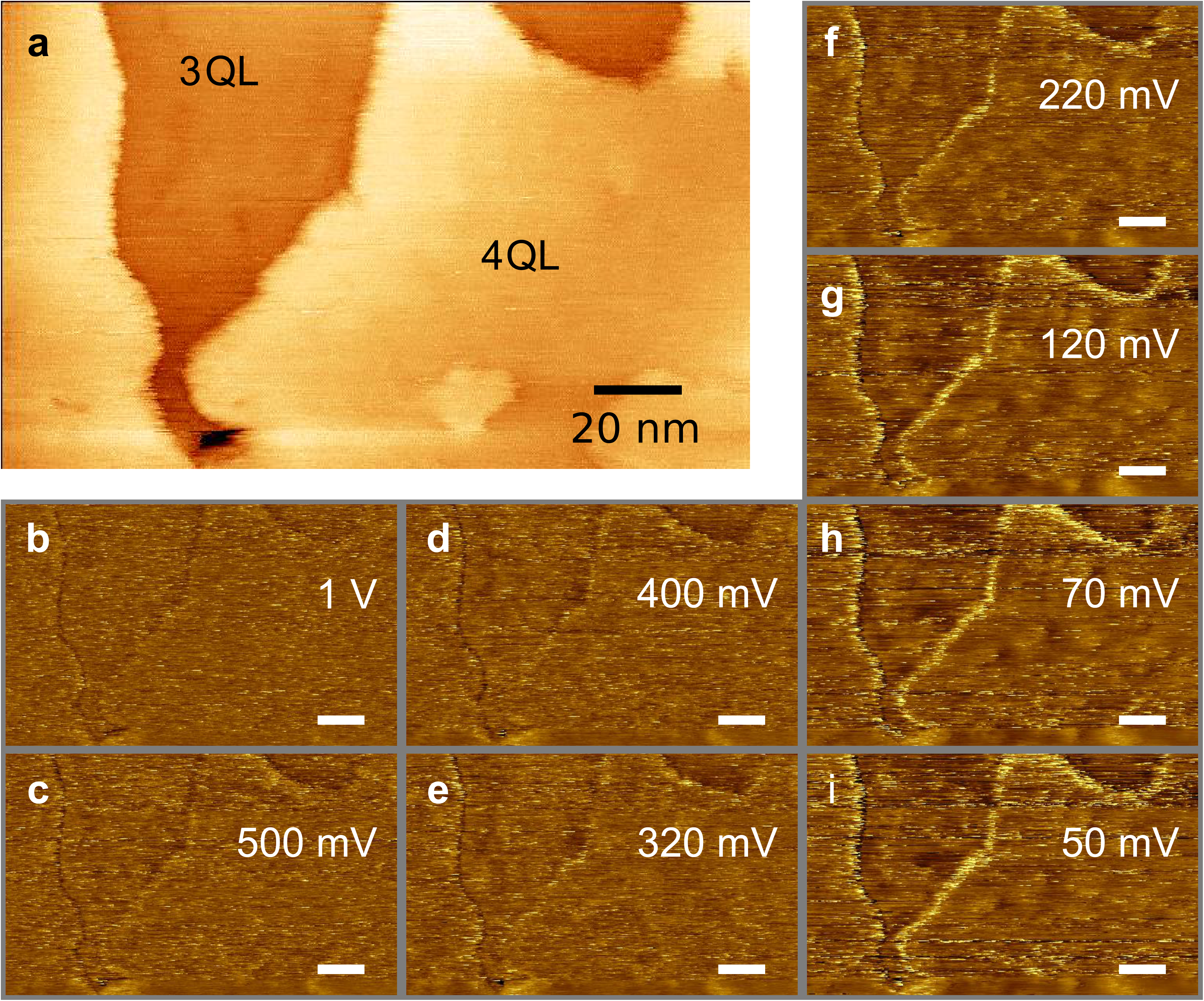}
		\caption{Differential conductance maps close to a step edge of the (Bi$_{0.16}$Sb$_{0.84})_2$Te$_3$ film. a) STM topography scan of the investigated area, including a step edge between a $3\,$QL terrace and a $4\,$QL terrace. b)-i) Corresponding spectroscopic $\mathrm{d}I/\mathrm{d}V$ maps, recorded in constant current mode at different sample bias voltages between $1\,$V and $50\,$mV. A pronounced feature in the $\mathrm{d}I/\mathrm{d}V$ signal along the step edge   at voltages below $220\,$mV indicates an increased local density of states at the step edge, thus an edge state.}
		\label{fig:fig3}
	\end{figure}

	Edge states typically show an increased LDOS which can be detected in STM and STS. For the example of topologically protected edge states this was demonstrated, e.g., in Refs. \onlinecite{Pauly2015, Reis2017, Luepke2020, Kim2016, Alpichshev2011}. Figure \ref{fig:fig3}a shows an STM topography image of a region close to a step edge between a 3\,QL and a 4\,QL film, and in Figure \ref{fig:fig3}b-i corresponding constant-energy $dI/dV$ maps are displayed. At energies larger than $220\,$meV, the maps convey a uniform density of states. Specifically, they exhibit no edge features. Yet, for energies below $220\,$meV a distinct LDOS signal along the step edges is observed. Since it occurs at the same magnitude regardless of whether the scanning tip ascends to the 4QL terrace or descends to the 3QL terrace (both types of tip trajectories are present in Fig.~\ref{fig:fig3}), we can rule out the possibility that the $dI/dV$ signal at the step edge is due to a feedback artifact.
	
	From angle-resolved photoemission spectroscopy it is known that the Fermi level at the top surface of the (Bi$_{0.16}$Sb$_{0.84})_2$Te$_3$ film is located $50\,$meV above the Dirac point and about $200\,$meV below the bulk conduction band \cite{Leis2021}. Therefore, the spectroscopic energy range of 0 $ \leq eV_\mathrm{b} \leq$ 200\,meV should correspond to states in the bulk band gap. Thus, we can conclude that the increased LDOS at the step edges in Fig.~\ref{fig:fig3}f-i must belong to states in the bulk band gap. The fact that these features disappear at a somewhat higher energy ($\sim 220$\,meV) than expected from the bulk band gap may be explained by a thickness-dependent band gap \cite{Jiang2012}.
	
	Characteristic LDOS signals at step edges as observed here are also found in other systems and have been interpreted as non-trivial QSH edge states \cite{Pauly2015, Reis2017, Luepke2020, Kim2016, Alpichshev2011}. Following this argumentation, we might conclude from Fig.~\ref{fig:fig3} that the 3\,QL (Bi$_{0.16}$Sb$_{0.84})_2$Te$_3$ film constitutes a QSH phase, while the 4\,QL film is topologically trivial (or vice versa), leading to a one-dimensional topological edge state at the boundary between the two. It must be noted, however, that from spectroscopic data alone it is not clear whether observed edge states are in fact topologically non-trivial. Therefore, it is necessary to explicitly prove the topological nature of  spectroscopically observed edge states. One way to achieve this is the measurement of the electrical conductance in these edge states. If an edge state is topological, then electrons injected into it should be protected from scattering and therefore travel ballistically along the edge, even if it is rugged \cite{Essert2015}. 
	This will lead to conductances that are substantially larger than the reference conductance in the two-dimensional interior (i.e., on the terraces) of the film.
	
	In the remainder of the paper, we will follow this approach, using a four-tip STM with which ballistic transport signatures in edge channels can in principle be detected by spatially resolved resistance measurements \cite{Picciotto2001, Baringhaus2014, Aprojanz2018APL, Aprojanz2018NatCom}. As mentioned above, in the present case the challenge is to precisely position our STM tips as close as possible to the rugged step edges of the MBE-grown wedge-like (Bi$_{0.16}$Sb$_{0.84})_2$Te$_3$ film. With the methodology described in the experimental section of this paper, we have been able to overcome this difficulty and have performed  distance-dependent four-point resistance measurements in close vicinity to the step edges.
	
	\subsection{Step-edge conductance by four-point resistance measurements along a step edge}
\begin{figure*}[t]
		\includegraphics[width=0.75\linewidth]{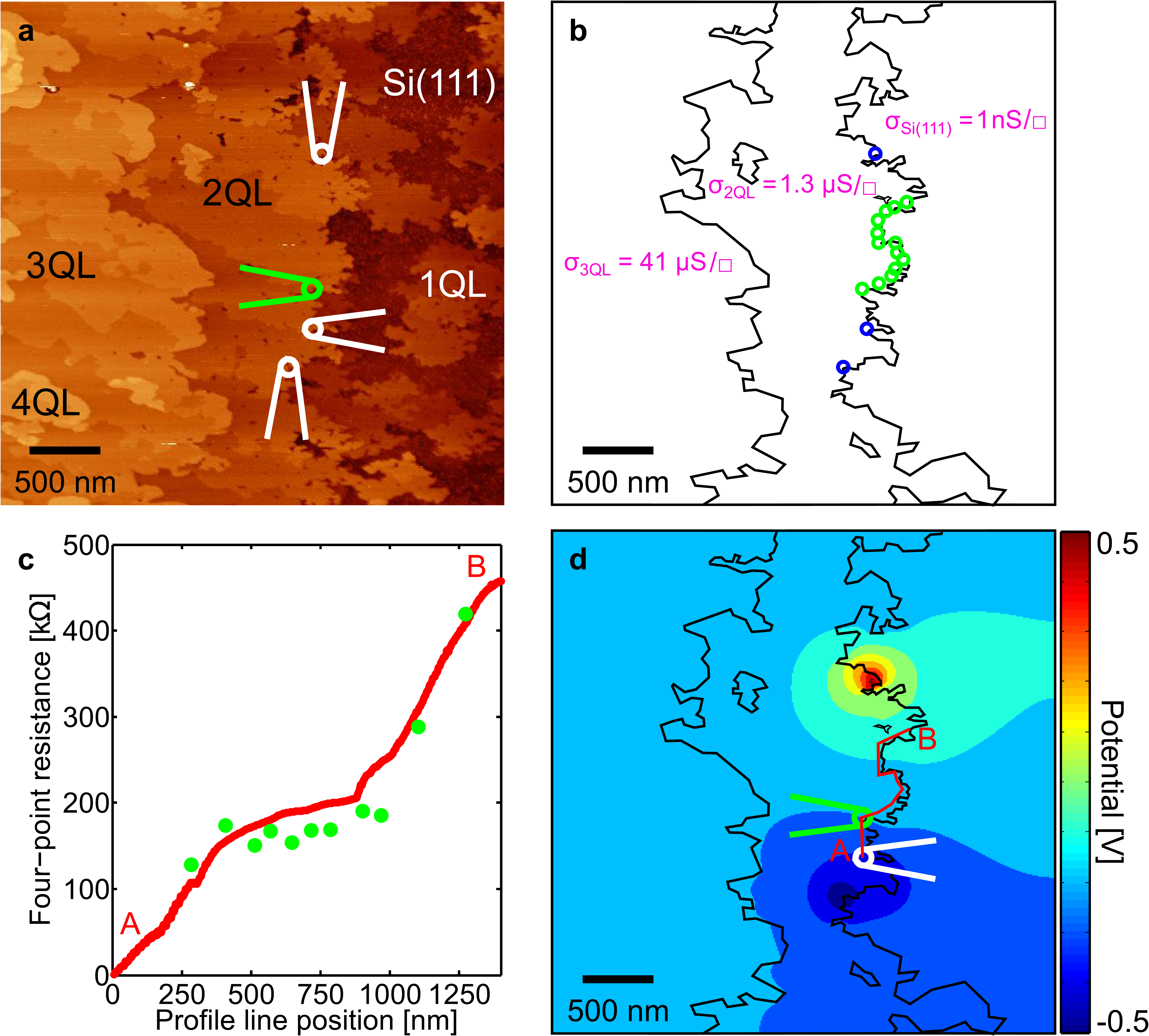}
		\caption{Four-point resistance measurements along a step edge.
			a) STM topography image of the investigated (Bi$_{0.16}$Sb$_{0.84})_2$Te$_3$ film. The image frame corresponds to the blue square in Fig.~\ref{fig:fig2}a.  White symbols indicate the positions of the current-injecting tips (top and bottom) and the fixed voltage-probing tip. The green symbol indicates the mobile voltage-probing tip. All tips are contacting the step edge between the 2\,QL terrace and the Si(111) substrate. b) Outline of the step edges with terrace sheet conductivities indicated \cite{Leis2021}. Fixed and mobile tip positions are indicated by blue and green open circles, respectively. c) Measured four-point resistance $R^\mathrm{4P}_\mathrm{2D}$ (green circles) vs.~position along the step edge between points A and B, exhibiting an approximately linear distance dependence ($\geq 100\,\mathrm{k\Omega}/\mathrm{\mu m}$), in comparison with the result of a finite-element simulation without a highly conductive edge channel (red curve). d) Color plot of the electric potential $\phi(x,y)$ from the finite-element simulation. The step edges are highlighted as black lines. The white tip symbol indicates the fixed voltage-probing tip at A, the red symbol the mobile voltage-probing tip at the first of the green data points in panel c. The trajectory of the mobile voltage-probing tip is indicated by the red line.}
		\label{fig:fig4}
	\end{figure*}

	Principally, there are several four-point measurement geometries in which ballistic edge channels  can be identified. In this section, we use a geometry in which all four tips are lined up along a step edge -- which does not need to be straight -- to search for a possible ballistic conductance channel at the boundary of a 2\,QL  (Bi$_{0.16}$Sb$_{0.84})_2$Te$_3$ film to the bare Si(111) substrate. The STM image in Fig.~\ref{fig:fig4}a shows the corresponding step edge and a representative measurement configuration of the four tips (white and green symbols). The precise sequence of measurement positions of the mobile voltage-probing tip is shown Fig.~\ref{fig:fig4}b (green open circles). In Fig.~\ref{fig:fig4}c, the measured four-point resistances are plotted versus the position of the mobile voltage-probing tip along the curved profile line (green dots). We observe a kink at the profile line position $\sim 1\,\mu$m, with different linear slopes to left and right of the kink. 
	
	Already a straightforward qualitative analysis reveals that the data in Fig.~\ref{fig:fig4}c are not consistent with the existence of a ballistic edge channel along the step edge. Since there can be no potential drop in a ballistic channel \cite{Datta1995}, its four-point resistance should be independent of the distance between the voltage probes.  This clearly is in contradiction with the measured data. Moreover, also the magnitude of the observed four-point resistance is at odds  with what would be anticipated for a ballistic conductance channel. For a single, i.e. not spin-degenerate, ballistic channel, one expects four-point resistances between $R_\mathrm{b}^\mathrm{4P} = 0$ and $R_\mathrm{b}^\mathrm{4P} = h/e^2 = 25.8\,$k$\Omega$ for the limiting cases of non-invasive and invasive voltage probes, respectively \cite{Picciotto2001, Aprojanz2018APL}. The significantly higher four-point resistances that we measure in our experiment ($\geq 100\,\mathrm{k\Omega}/\mathrm{\mu m}$) strongly suggest that there is no ballistic conductance channel at the step edge in Fig.~\ref{fig:fig4}a.
	
	It should be noted, however, that a QSH phase could in principle also exhibit a behavior that is devoid of ballistic transport signatures. This would be the case if the Fermi level were not in the band gap or if inelastic scattering were present. In the first case, interband scattering could occur between the one-dimensional edge channel and the bulk states in the two-dimensional interior, as shown in Fig. 1c. As a result, the conductance in the edge channel would be reduced. Because the Fermi level position in the bands of the two-dimensional film interior  would at the same time increase the two-dimensional sheet conductivity, an existing ballistic conductance channel at the edge could in fact be masked in our transport experiments, and with it a QSH phase in the film.  However, because of the large surface band gap $\Delta_{2\,\mathrm{QL}} \approx 250\,$meV \cite{Jiang2012, Foerster2016} and the Fermi level position $E_\mathrm{F} \approx 50\,$meV above the Dirac point \cite{Leis2021} (with a spread of about 25\,meV at room temperature), inter-band scattering seems very unlikely here. Regarding the second case, it should be noted that inelastic scattering, e.g. by electron-phonon interaction, is possible in principle, even if elastic backscattering of the spin-polarised QSH edge states is excluded by time-reversal symmetry. Obviously, inelastic scattering could increase the resistance of the edge channel beyond what is expected for the ballistic case. The strength of this effect depends on the particular system, but has not yet been quantified at conditions close to room temperature. Usually, the opposite behaviour is observed, namely that the resistance decreases as a function of temperature due to thermal activation of transport in two-dimensional bulk states \cite{Wu2018,Dolcetto2016}. However, in our experiment, bulk transport is strongly suppressed at room temperature (in particular for the 2\,QL terrace), which imposes stringent conditions on the inelastic scattering rate that would be required to obtain a QSH edge state with a resistivity of up to $\geq 100\,\mathrm{k\Omega}/\mathrm{\mu m}$. Therefore, the absence of a QSH edge state remains the most convincing explanation for the experimental data.
	
	Having all but ruled out the presence of ballistic conductance channels at the edge of the 2\,QL film, we still need to explain the distinctive behavior of the four-point resistance in Fig.~\ref{fig:fig4}c. To this end, we performed a finite-element simulation of a classical boundary value problem, taking into account the precise shape of the terraces from Fig.~\ref{fig:fig4}a, the exact positions of the current injection points (at which the potential in the calculation was set to $+0.5\,$V and $-0.5\,$V), and the known sheet conductivities \cite{Leis2021} of both the 2\,QL (Bi$_{0.16}$Sb$_{0.84})_2$Te$_3$ film and the Si(111) substrate, $\sigma_{2\,\mathrm{QL}}=1.3\,\mu$S$/\square$ and $\sigma_{\mathrm{Si(111)}}=1$\,nS$/\square$. The resultant distribution of the electric potential $\phi$ in the film surface is displayed in Fig.~\ref{fig:fig4}d. From this calculated potential, we extracted the four-point resistance $ R^{\mathrm{4P}}_{\mathrm{2D}}=\delta \phi/I$ measured between the fixed and the mobile voltage-probing tips at different positions along the step edge (red line in Fig.~\ref{fig:fig4}d and red curve in Fig.~\ref{fig:fig4}c), where $\delta \phi$ is the calculated potential difference between points at which voltage-probing tips are placed and $I$ is the current. We note the excellent agreement of the experimental and simulated four-point resistances (without any scaling factor). Even the kink in the four-point resistance around 1\,$\mu$m along the step edge is faithfully reproduced in the simulation. This implies that this feature originates from the specific shape of the step edge, and does not require any change in local conductivities within the film for its explanation, as would be instigated, e.g., by a ballistic conductance channel at the edge of a QSH phase. All the evidence put together, we thus unequivocally conclude that the present 2\,QL film of (Bi$_{0.16}$Sb$_{0.84}$)$_2$Te$_3$ does not exhibit ballistic edge channels and, since the neighboring Si substrate is definitely topologically trivial, also does not represent a QSH phase either.
	
	An analysis of the band structures of both Bi$_2$Te$_3$ and Sb$_2$Te$_3$ in the thin-film limit with many-body perturbation theory in the so-called GW approximation has revealed that 2\,QL films of Bi$_2$Te$_3$ are expected to be a QSH insulator, with a band gap of 0.15\,eV \cite{Foerster2016}. With a gap of this size, the edge channels should be observable in  experiment, if indeed the Fermi level is in the 2D bulk band gap. The fact that we do not observe this QSH phase in our transport experiment does not necessarily contradict this result, because in the same calculation 2\,QL films of Sb$_2$Te$_3$ are found to be topologically trivial \cite{Foerster2016}, and it is not a priori clear how the mixed compound (Bi$_{0.16}$Sb$_{0.84}$)$_2$Te$_3$ would behave, although one might expect it to resemble Sb$_2$Te$_3$ more closely than Bi$_2$Te$_3$. Therefore, we have also carried out nanoscale four-point transport experiments on 2\,QL films of Bi$_2$Te$_3$ in order to search for a ballistic conductance channel along the edges. Because of the intrinsic $n$ doping of this material, however, this requires to shift the Fermi level from the conduction band into the band gap. Since our samples are MBE-grown on silicon on insulator substrates, this can principally be achieved by gating \cite{Leis2021}. Unfortunately, however, no change of the sheet conductivity upon gating up to 200\,V was observed in initial experiments, showing that our gating capability is not sufficient to move the Fermi level into the band gap of Bi$_2$Te$_3$. As a result, the terrace conductivity of this sample was significantly higher ($\sim 200\,\mu $S$/\square$) than that of (Bi$_{0.16}$Sb$_{0.84}$)$_2$Te$_3$, ($1.3\,\mu $S$/\square$), such that we were unable to distinguish any edge conduction. However, this does not mean that the Bi$_2$Te$_3$ films are topologically trivial. It just illustrates our inability to suppress their two-dimensional sheet conductivities sufficiently.

	\subsection{Step-edge conductance by four-point resistance measurements across a step edge}
	Although the results of the analysis in the previous section are unambiguous, one may ask how critical it is to bring the tips as close as possible to the step edge (and thus also to the edge state) in order not to miss the ballistic edge channel whose width is difficult to gauge. In this section, we therefore explore whether a tip configuration with less stringent requirements can be used to identify the presence or absence of a ballistic edge channel as well. Specifically, we position three tips modestly close and roughly parallel to a step edge, while the fourth, i.e., one of the voltage-probing tips, is moved through a sequence of measurement positions along a line that runs approximately perpendicular to the step edge. We apply this measurement geometry to the same sample as before, but we focus on the step edge between the 2\,QL and 3\,QL films. Since we have already established that the 2\,QL (Bi$_{0.16}$Sb$_{0.84})_2$Te$_3$ film is trivial, we can thus analyze the topological properties of the 3\,QL film. 
	
\begin{figure*}[t]
		\includegraphics[width=0.7\linewidth]{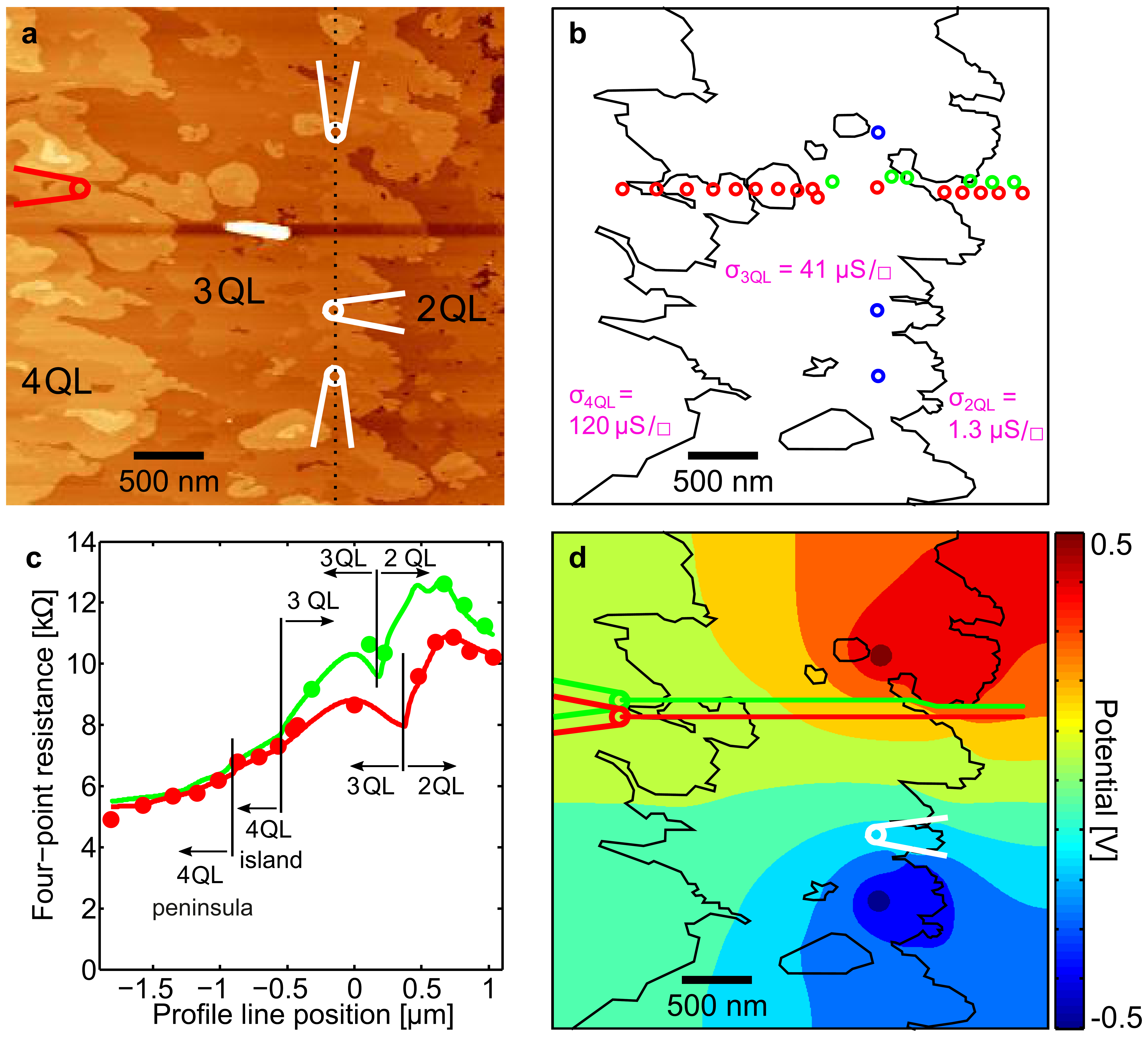}
		\caption{ Four-point resistance measurements perpendicular to a step edge. a) STM topography image of the investigated (Bi$_{0.16}$Sb$_{0.84})_2$Te$_3$ film. White symbols indicate the positions of the current-injecting tips (top and bottom) and the fixed voltage-probing tip. The red tip symbol indicates the mobile voltage-probing tip. While the static STM tips are placed along the step edge between the 3\,QL and 2\,QL terraces, the mobile voltage-probing tip is moved perpendicularly to the step edge.(b) Outline of the step edges with terrace sheet conductivities indicated \cite{Leis2021}. Fixed tip positions are indicated by blue, mobile tip positions by red and green open circles.
			(c) Measured four-point resistance $R^\mathrm{4P}_\mathrm{2D}$ (red and green circles) vs.~position, in comparison with the result of a finite-element simulation without a highly conductive edge channels (red and green curves). Characteristic positions along the profile line are marked. 
			(d) Color plot of the electric potential $\phi(x,y)$ from the finite-element simulation without highly conductive channel. The step edges are highlighted as black lines. The white symbol indicates the fixed voltage-probing tip, the green/red symbol the mobile voltage-probing tip at the first of the red data points in panel c. The trajectories of the mobile voltage-probing tips is indicated by the green/red lines.}
		\label{fig:fig5}
	\end{figure*}

	Figure~\ref{fig:fig5}a shows an STM image of the boundary region between 2\,QL and 3\,QL terraces of the sample. The two current-injecting tips and the immobile voltage-probing tip are placed on the 3\,QL terrace along the vertical dotted line, at a distance of $\sim 200\,$nm from the step edge to the 2\,QL terrace. The measurement positions of the mobile tip are indicated by red and green symbols in Fig.~\ref{fig:fig5}b for two measurement series, which are slightly offset against each other. As before, we simulate the transport experiment by solving the corresponding boundary value problem with the known sheet conductivities. 
	%$\sigma_{3\,\mathrm{QL}}=41$\,$\mu$S$/\square$ and $\sigma_{4\,\mathrm{QL}}=120$\,$\mu$S$/\square$. 
	The red and green curves in Fig.~\ref{fig:fig5}c show the calculated four-point resistances along the two horizontal lines in Fig.~\ref{fig:fig5}d. The agreement between the measured four-point resistances and the finite-element simulation is again exceptionally good, including details such as the maxima at $\sim 0.5\,\mu$m, the dip in the green data set at $\sim 0.15\,\mu$m, and the vertical offset between red and green data sets. 
	
	Whether the presence of a ballistic edge channel will modify the measured four-point resistances depends on the ratio of the terrace sheet conductivity $\sigma_\mathrm{2D}$ to the total contact resistance $R_\mathrm{b} = h/e^2 = 25.8\,$k$\Omega$ involved in entering and leaving the ballistic conductance channel. If $1/\sigma_\mathrm{2D} \gg R_\mathrm{b}$, the ballistic channel acts as a perfect conductor. Whenever this strong condition applies, the current path through the ballistic channel is strongly preferred over the current path through the plane, resulting in a significantly modified four-point resistance. In the opposite case the contact resistance $R_\mathrm{b}$ will impede the current from entering the ballistic channel.
	
	In the present case the above-mentioned strong condition is not fulfilled, since $1/\sigma_\mathrm{3\,QL} \approx 24\,\mathrm{k \Omega}/\square$ is of the same order as $R_\mathrm{b}$. However, there is a weaker condition which indicates that a significant part of the current flows through the  ballistic channel. To illustrate this, consider a current injection into the plane at positions close to the potential ballistic edge channel (as in Fig.~\ref{fig:fig5}a). In this case, the currents through the ballistic channel and the plane can be approximated by a parallel resistor model in which the resistance through the plane is given by its two-point resistance $R_\mathrm{2D}^\mathrm{2P}$. If the weaker condition $R_\mathrm{2D}^\mathrm{2P} > R_\mathrm{b}$ is fulfilled, the parallel resistor model tells us that a large part of the current flows through the ballistic channel, which influences in turn the measured four-point resistances.
	
	The two-point resistance in a plane is given by \cite{Voigtlaender2018}
	\begin{equation}
		R^\mathrm{2P}_\mathrm{2D} = \frac{1}{\pi \sigma_{\mathrm{2D}}} \ln \left( \frac{\Delta x - R}{R} \right),
		\label{eq:2pointR}
	\end{equation}
	\newline
	where $\Delta x$ is the distance between the current-injecting tips and $R$ their contact radius. Considering the case in Fig.~\ref{fig:fig5}a with $\Delta x \approx 1.7\,\mu\mathrm{m}$ and $R \approx 10\,\mathrm{nm}$, the two-point resistance on the 3\,QL terrace results as $R^\mathrm{2P}_\mathrm{2D}  \approx 40\,\mathrm{k}\Omega$. The actual two-point resistance is expected to be nearly twice this value, if---as in the present case---the tips are close to a terrace of low conductivity and thus only one half-plane contributes to the two-point sheet conductance. Thus, in our case the condition $R^\mathrm{2P}_\mathrm{2D} \approx 80\,\mathrm{k}\Omega  > R_\mathrm{b} = 25.8\,$k$\Omega$ is fulfilled and a significant modification of the measured four-point resistance is expected if a ballistic channel was present. The experimental data are therefore not consistent with the presence of a highly conductive edge channel at the step edge between the 2\,QL and the 3\,QL films. We note that according to the above argument, the identification of one-dimensional ballistic conductance channels is easier if $\sigma_\mathrm{2D}$ is small---for this reason we have deliberately chosen a (Bi$_{x}$Sb$_{1-x})_2$Te$_3$ compound that has its Fermi level in the bulk band gap, such that the charge carrier density and hence the background conductivity on the terraces is as low as possible. 
	
	However, a note of caution is in order: the bulk band gap  of the 3\,QL film ($\Delta_{3\,\mathrm{QL}} \approx 60\,$meV) is much smaller than that of the 2\,QL film \cite{Jiang2012, Foerster2016}. Because the estimated Fermi level position of $E_\mathrm{F} \approx \pm 50\,$meV with respect to the Dirac point \cite{Leis2021} is in the same range as this band gap, inter-band scattering in between the ballistic edge channel and the two-dimensional interior of the film cannot be excluded with certainty. If the Fermi level was indeed located in the conduction band, the inter-band scattering would reduce the edge state conductance; at the same time, the sheet conductivity of the two-dimensional interior of the film would increase, making it difficult to identify signatures of the ballistic transport in our transport data. For the 3\,QL  (Bi$_{0.16}$Sb$_{0.84})_2$Te$_3$ film we can therefore conclude that, if the Fermi level is in gap, the film must be topologically trivial. If, however, the Fermi level is located in the band, no definite conclusion regarding the existence of a QSH phase can be reached.

	\subsection{Analytical model of the four-point resistance along an edge between half-planes with distinct sheet conductivities}
	The excellent agreement between the experimental data and the simulation in Fig.~\ref{fig:fig4}c and Fig.~\ref{fig:fig5}c encourages a further analysis of specific features in the four-point resistance profiles. To this end, we calculated such profiles analytically for two generic situations, with and without the presence of a highly conductive edge channel. As we will see in this section, the four-point resistances in the two limiting situations differ strongly. In many cases, these qualitative differences will allow the classification of films as topologically non-trivial by a straightforward visual inspection of the measured distance-dependent four-point resistance profiles. We note that the calculations in this section presume the condition  $1/\sigma_\mathrm{2D} \gg R_\mathrm{b}$.

\begin{figure*}
		\includegraphics[width=\linewidth]{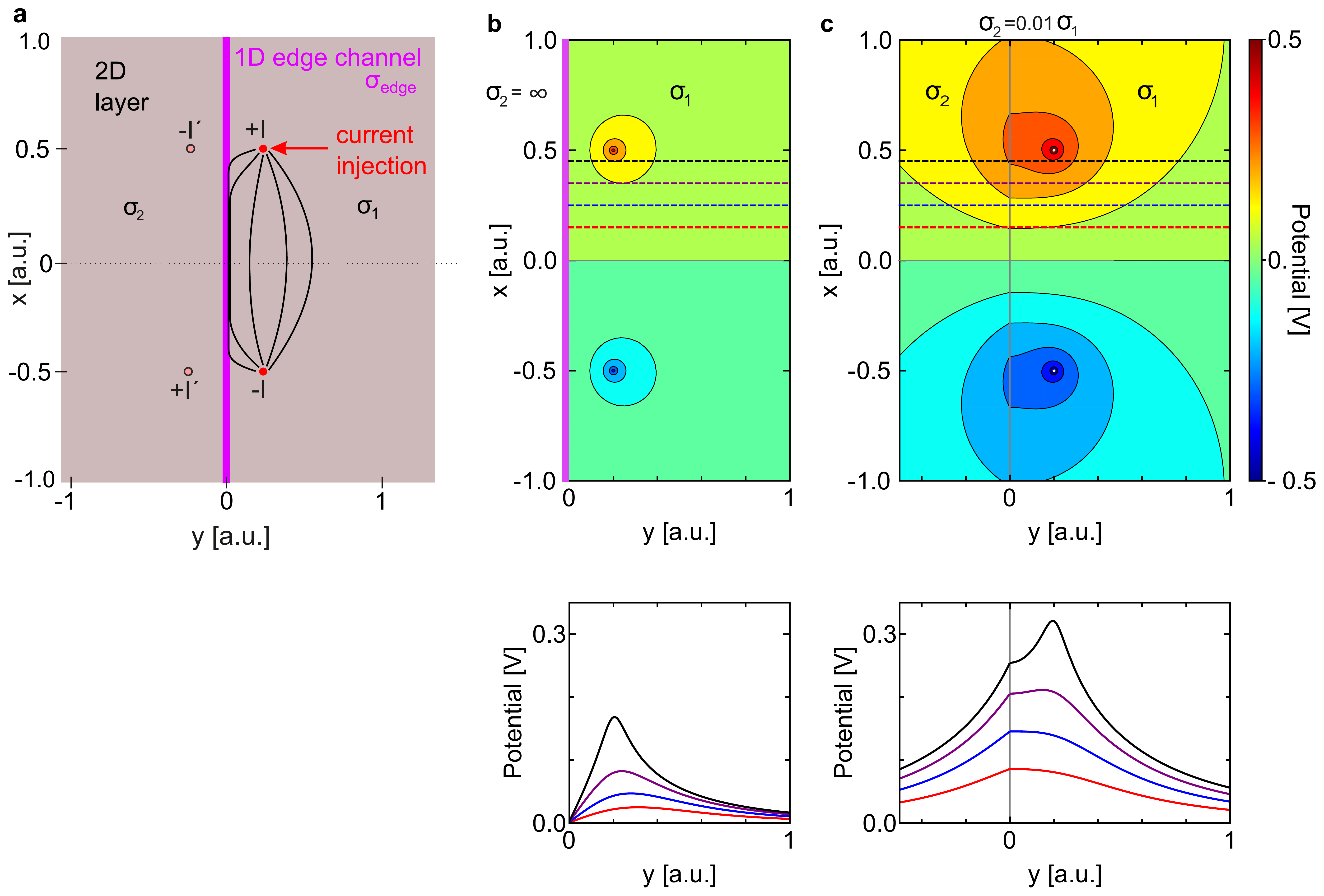}
		\caption{Analytical solution for the electric potential $\phi(x,y)$ close to the edge between two half-planes. a) Geometry of the boundary value problem. Two half-planes with sheet conductivites $\sigma_1$ (for $y>0$) and $\sigma_2$ (for $y<0$) meet at $y=0$. A current $+I$ is injected at $(x_0,y_0)$ and withdrawn ($-I$) at $(-x_0,y_0)$. If $\sigma_2\rightarrow \infty$, this geometry emulates the existence of a one-dimensional ballistic conductance channel at $y=0$ (magenta line). Positions of image current sources are  indicated by $I'$ (see methods section for more details). b) Color map of $\phi(x,y)$ for $\sigma_2\rightarrow \infty$. For $y<0$, the potential is zero. Profiles for $x=\mathrm{const.}$ are shown at the bottom of the panel: for $y\rightarrow 0$, the potential increases, reaches a maximum and falls to zero at the ballistic conductance channel.  
			c) Color map and profiles of $\phi(x,y)$ for $\sigma_2=0.01\sigma_1$, emulating the case of a step in the two-dimensional sheet conductivity at $y=0$. The potential tends to increase towards this step, indicating the existence of a resistance barrier there. }
		\label{fig:fig6}
	\end{figure*}

	To calculate the four point resistance $ R^{\mathrm{4P}}_{\mathrm{2D}}=\delta \phi/I$, we analytically determined the potential distribution $\phi (x,y)$ due to the stationary current distribution, from which the potential difference $\delta \phi$ between arbitrary positions of the two voltage-probing tips was obtained. We considered the following geometry (Fig.~\ref{fig:fig6}a): The two current-injecting tips were placed at $(-x_0,y_0)$ and $(x_0,y_0)$. We assumed constant sheet conductivities $\sigma_{1}$ for $y>0$ and $\sigma_{2}$ for $y<0$ (for simplicity, we leave out the subscript `2D' from now on). Applied to the present problem, this corresponds to a step between two terraces of different heights at $y=0$. We further assumed that $\sigma_1>\sigma_2$ (for a stepped TI film this would mean that the film thickness for $y>0$ is higher). Within this model, we considered two cases: If the step at $y=0$ is topologically trivial, then the system is fully described by the two finite sheet conductivities. If on the other hand one of the films is in a QSH state, a ballistic conductance channel will run along the edge at $y=0$; we accommodated this case into our model by letting $\sigma_2\rightarrow \infty$ and restricting our solution to the half-plane $y>0$. 
	
	The calculated potentials (see section theoretical methods for the derivation of the corresponding equations) are plotted in Fig.~\ref{fig:fig6}b and c for two cases, one of them modelling the presence of a ballistic edge channel %(Eqs.~\ref{phi1limit}, \ref{phi2limit} and (
	Fig.~\ref{fig:fig6}b), the other its absence 
	%(Eqs.~\ref{eq:phi1}, \ref{eq:phi2} and 
	(Fig.~\ref{fig:fig6}c). Turning to Fig.~\ref{fig:fig6}b first, we observe that  the potential is distorted from the simple dipolar distribution that would be observed if the current was injected into an infinite two-dimensional plane. The profiles perpendicular to the edge which are displayed in the bottom part of Fig.~\ref{fig:fig6}b reveal a strong drop of the potential towards the edge and in particular a near-constant potential $\phi_1(x,y)\approx 0$ along the edge for $y\rightarrow 0$, to achieve continuity with $\phi_2(x,y)=0$ for $y<0$. Since constant potential indicates low resistance and thus large current densities, Fig.~\ref{fig:fig6}b reveals that the injected current preferably flows through the ballistic edge channel. 
	
	In contrast, the situation without the ballistic edge channel, but with a poorly conducting half-plane $y<0$, leads to a very different distribution of the potential, as shown in Fig.~\ref{fig:fig6}c for the specific case $\sigma_2=0.01\sigma_1$. 
	In Fig.~\ref{fig:fig6}c, the strong variation with $x$ of the potential $\phi_1(x,y)$ for $y\rightarrow 0$ indicates a large resistance along the edge. As a consequence, the current is not only inhibited from entering the poorly conducting half-plane, but is suppressed already for $y>0$ while $y$ approaches zero from above. We note that the $\phi_2$ drops again for increasingly negative $y$ away from the edge. However, the `resistance barrier' on both sides of the edge means that the actual current density there will be very small. The potential $\phi_1(y\rightarrow 0)$ does not vanish, as it was in the case for a perfectly conducting channel at $y=0$.
	
	Comparing the two situations in Fig.~\ref{fig:fig6}b and c, it is clear that all experimental data reported in this paper resemble the situation in Fig.~\ref{fig:fig6}c. On the one hand, this confirms that no indications of a ballistic edge channel are present for (Bi$_{0.16}$Sb$_{0.84})_2$Te$_3$. On the other hand, this also shows that the behavior of the analytical solution of a straight step edge
	%in Eqs.~\ref{eq:phi1} and~\ref{eq:phi2} 
	goes a long way to providing a qualitative guide to the existence or non-existence of a ballistic edge channel and therefore a QSH phase, even in cases when the actual step edge is far from straight. We note in this context that for strongly defective step edges, the edge states of a QSH insulator are expected to retract away from the step edge into the two-dimensional interior, where the ballistic channel may preserve a larger degree of straightness \cite{Essert2015}. 
	
	\section*{Conclusion}
	Employing the Jülich multi-tip STM, we performed nanoscale charge transport measurements in the vicinity of single quintuple-layer steps of TI films. Due to the precise positioning and navigation capabilities of our STM tips down to the nanoscale, we were able to spatially resolve four-point resistances at the TI film. In particular, we successfully overcame the challenge of the rather rough step edges in our MBE-grown samples. Our local conductance data proved to be informative with respect to the presence or absence of ballistic edge channels that would indicate a QSH phase. In this sense, we achieved our goal to close the research gap between the spectroscopic characterization with scanning probe methods at the nanometer scale on the one hand, and conventional transport experiments at the micrometer scale on the other, even in the face of a non-ideal nanoscale sample geometry (rugged edges). 
	
	From our experimental data, we could conclude that no ballistic edge channels were present in our sample for 2\,QL and 3\,QL films, despite a gap opening in the TSS of the underlying three-dimensional TI, which manifested itself in an exponentially decreasing sheet conductivity for films with thicknesses below 5\,QL and despite an edge state signature in LDOS. Instead, we found that the measured four-point resistances agree \textit{quantitatively} with results from finite-element simulations based on, first, the actually observed intricate nanoscale sample geometry and, second, measured conductivities on the TI terraces. We also found \textit{qualitative} agreement with an analytical solution of the Laplace equation in the idealized geometry of a straight edge at which no ballistic conductance channel is present. 
	
	The absence of ballistic edge channels implies a topologically trivial phase for a film thickness of $2\,$QL, for which $E_\mathrm{F}$ is clearly located in the band gap. In contrast, for the film thickness of $3\,$QL, the absence of ballistic edge channels principally allows two interpretations: first, the absence of a QSH phase, or second, a QSH phase with topologically protected edge channels while $E_\mathrm{F}$ is located in the two-dimensional bulk band. The latter situation would result in an enhanced terrace conductivity, while at the same time enabling scattering from the edge channel into bulk bands (see Fig. \ref{fig:fig1}c), thereby undermining any ballistic transport  along the edges. We note, however, that a Fermi energy $E_\mathrm{F}$ in the bulk band appears unlikely, because the terrace conductivity was found to be rather small for $3\,$QL films. Thus, we interpret the increased local density of states observed in $dI/dV$ maps at the step edges as originating from trivial edge states with a conductivity that is not higher than that of the surrounding terraces. In summary, we conclude that the investigated compound (Bi$_{0.16}$Sb$_{0.84}$)$_2$Te$_3$ does not exhibit a QSH phase for 2 to 5\,QL films, because  ballistic edge states, being the hallmark of a QSH phase, are not present. This is in agreement with theoretical predictions for the binary compound Bi$_2$Te$_3$ that is close to the stoichiometry of our present sample \cite{Foerster2016}. 
	
	On a more general note, we were able to demonstrate that STM-based multi-tip transport experiments are a powerful and generic method to search for ballistic edge channels. But we have also seen that firm conclusions could only be drawn with the proviso that the Fermi level is located in the bulk band gap. This touches upon a problem of transport measurement in general, regardless of the methodology: if the bulk conductivity is too large, then it becomes difficult or impossible to distinguish the ballistic transport in the one-dimensional edge channels from the transport in the two-dimensional film interior. In this respect, the MBE-grown (Bi$_x$Sb$_{1-x}$)$_2$Te$_3$ films turned out to be problematic, because they require a certain stoichiometry to put the Fermi level into the bulk gap, but at the same time this stoichiometry influences the topological properties. In the near future, we therefore plan to apply the methodology of the present paper to exfoliated films, for which unintentional doping is expected to be less of a problem. Promising materials in this context are magnetic topological insulators, which exhibit a quantum anomalous Hall state, and van der Waals materials, such as the QSH insulator WTe$_2$, which allow an effective gate tuning.
	Moreover, low temperature measurements could also reveal the impact of inelastic scattering on the transport properties of the edge channels.
	Finally we note that our technique is also compatible with the measurement of spin-polarized transport, if magnetic tips are used \cite{Leis2020}. This should finally bring us closer to the long-term goal of spin-polarized measurement of ballistic conductance channels with non-invasive contacts that can be freely positioned on the nanoscale.

	\subsection*{Acknowledgments}
	This work was supported by the German excellence cluster ML4Q (Matter and Light for Quantum Computing). Furthermore, the authors acknowledge the financial support by the Bavarian Ministry of Economic Affairs, Regional Development and Energy within Bavaria’s High-Tech Agenda Project ``Bausteine für das Quantencomputing auf Basis topologischer Materialien mit experimentellen und theoretischen Ans\"atzen'' (grant allocation no.\ 07 02/686 58/1/21 1/22 2/23).
	F.L. acknowledges funding by the Deutsche Forschungsgemeinschaft (DFG, German Research Foundation) within the Priority Programme SPP 2244 (project no. 443416235). 
	F.S.T. acknowledges support of the Deutsche Forschungsgemeinschaft through the SFB 1083, project A12. We  would like to thank R. Greven of ZEA-1 for the transfer of the STM data to the finite element program.
	%\end{acknowledgments}
	%
	%\section*{Author contributions}
	%F.S.T., B.V. and A.L. conceived the research. A.L. performed the experiments. A.L., V.C. and B.V. designed the experiments. The manuscript was written by A.L. All authors discussed and commented on the manuscript.
	
	%\section*{Additional Information}
	
	\subsection*{Conflict of interests}
	
	The authors declare no conflict of interest.

	\subsection*{Data availability}
	
	Data within the manuscript is available from the corresponding author upon reasonable request.
	
	\newpage 
	
\end{document}